\documentclass[journal,9pt]{IEEEtran}
\usepackage{graphicx,bm}
\usepackage{amsmath,amsfonts}
\usepackage{booktabs}

\graphicspath{{Figures/}}

\usepackage{epstopdf}
\usepackage{dcolumn}
\usepackage{bm}
\usepackage{lipsum}
\usepackage{multirow}
\usepackage{color}
\usepackage[dvipsnames]{xcolor}

\usepackage{amsmath}
\usepackage{amssymb}
\usepackage{mathrsfs}
\usepackage{turnstile}
\usepackage{graphicx}
\usepackage{color}
\usepackage{tensor}
\usepackage{bm}
\usepackage{cite}

\usepackage{epstopdf}

\usepackage{xcolor}
\usepackage{balance}
\usepackage{amsmath}

\ifCLASSOPTIONcompsoc
 \usepackage[caption=false,font=normalsize,labelfont=sf,textfont=sf]{subfig}
\else
 \usepackage[caption=false,font=footnotesize]{subfig}
\fi

\begin{document}

\title{Leaky-Mode Generation by Means of All-Metal \\ Corrugated Slotted Parallel-Plate Waveguides}

\author{Yuhuan Tong~\IEEEmembership{Student Member,~IEEE,}
Beatrice Ambrogi~\IEEEmembership{Student Member,~IEEE,} \\
Davide Comite~\IEEEmembership{Senior Member,~IEEE,} 
and 
Guido Valerio~\IEEEmembership{Senior Member,~IEEE}

\thanks{This work is supported by the Air Force Office of Scientific Research under award number FA8655-23-1-7018
and by the Horizon Europe Research and Innovation Program under the GENIUS Project, Marie Skłodowska-Curie Grant under Agreement 101072560.}
\thanks{Y. Tong and G. Valerio are with with Sorbonne Université, CNRS, Laboratoire de Génie Electrique et Electronique de Paris, 75252, Paris, France and with Université Paris-Saclay, CentraleSupélec, CNRS, Laboratoire de Génie Electrique et Electronique de Paris, 91192, Gif-sur-Yvette, France, e-mail:\{yuhuan.tong, guido.valerio\}@sorbonne-universite.fr}
\thanks{B. Ambrogi, and D. Comite are with the Department of Information Engineering, Electronics and Telecommunications, Sapienza University of Rome, Rome, Italy, e-mail:\{beatrice, comite\}@uniroma1.it}
\thanks{Manuscript received Month XX, 20ZZ.}}

\markboth{IEEE Transactions on Antennas and Propagation,~Vol.~XX, No.~YY, Month~20ZZ}%
{Tong \MakeLowercase{\textit{et al.}}: 2D-Corrugated Slotted Parallel-Plate Waveguides}


\maketitle

\begin{abstract}
Periodic leaky-wave antennas can be realized with planar low-profile structures supporting the generation of highly directional beams with frequency scanning from backward to forward directions. We propose a method to generate leaky modes with continuous scanning with an all-metal periodic 2-D (i.e., invariant along a direction normal to the propagation) waveguide. Periodic corrugations are introduced to support a slow wave, which is transformed with periodic slots etched on the top plate into a fast radiating waves. Suitable choices of the corrugation and the slot periods lead to a single fast backward-forward spatial harmonic being responsible for well-defined and directional radiation. This new method is demonstrated here proposing an original unit cell, properly optimized to suppress the open stop band. The analysis and the design is accomplished with a rigorous in house periodic method-of-moment code, allowing for the computation of complex modes in open 2-D waveguides. The leaky-wave antenna is studied and validated by means of full-wave simulations in Ka band. The structure features continuous beam scanning with broadside radiation at 29.7 GHz with beam scanning, realized gain, and efficiency comparable to different antenna technologies having thickness ten times larger.

\end{abstract}

\vskip0.5\baselineskip

\section{Introduction}

\IEEEPARstart{N}{ext} generation of wireless and navigation technologies are undergoing a significant spectral transition. The increasing congestion in the C and Ku bands is driving industrial applications toward K and Ka-bands, motivating renewed scientific interest in these frequency ranges. This transition will allow to meet the growing demand for higher data rates, to improve spectral efficiency, as well as to decrease device size \cite{lenses_comm}. In order to avoid the power consumption of complex beamforming networks and reduce material losses, all-metal leaky-wave antennas (LWAs) offer low profile designs capable to provide directive radiation and beam steering, due to their inherent frequency scanning capabilities \cite{jackson2012leaky}. Fast scanning LWAs can be ideal candidates for next-generation indoor positioning systems as well as for medium- and long-range tracking applications \cite{DirFindingJose}, while broadside radiation is also possible with LWAs, when optimized to maximize gain and bandwidth \cite{jackson2012leaky} for terrestrial and satellite communications. 
\begin{figure}[t]
	\centering
    \includegraphics[width=0.35\textwidth]{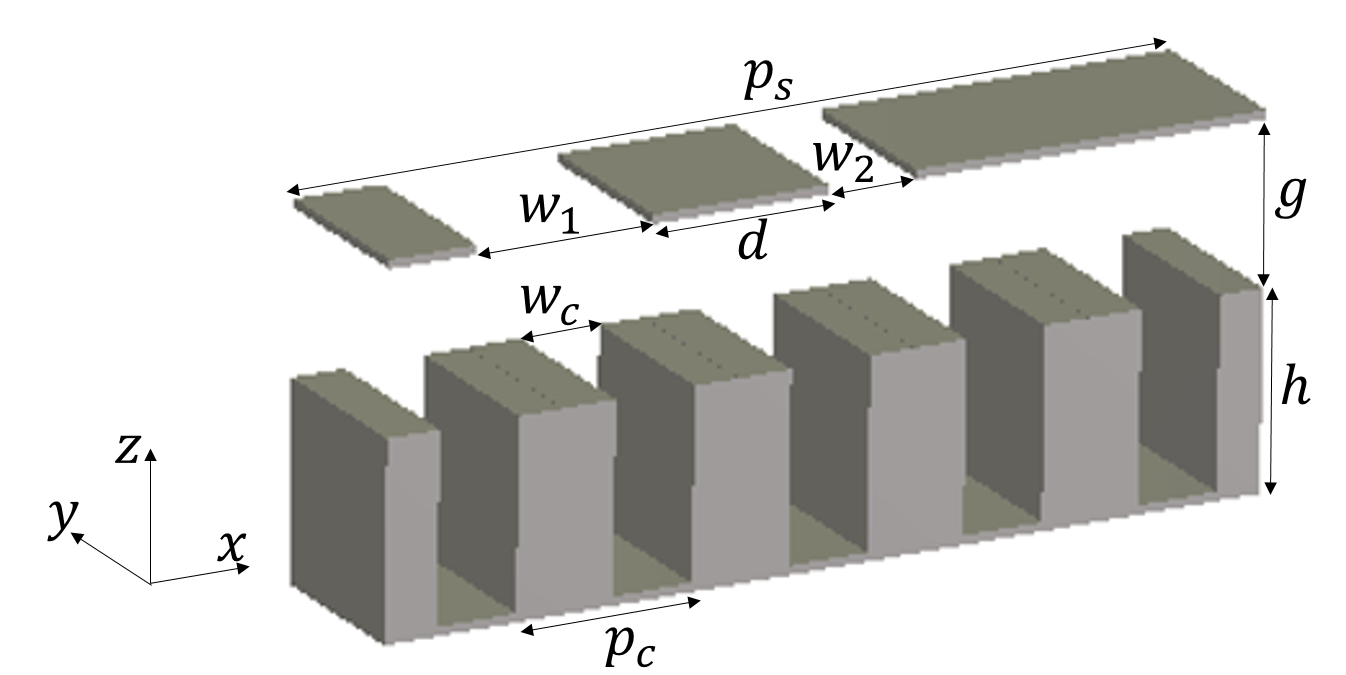} 
\caption{Three-dimensional view  of the metallic corrugated periodic antenna  under analysis, invariant along the $y$ direction, with relevant geometrical parameters.} 
\label{fig:geometry}
\end{figure}

The above-mentioned applications require a control of frequency dispersion, which can be obtained by including in the LWA artificial surfaces or periodic elements, such as periodic corrugations. Furthermore, if a backward-to-forward continuous scanning is expected, suitable techniques should be employed to suppress the LWA Open Stop-Band (OSB), a phenomenon resulting in a gain drop when the beam is directed toward broadside\cite{paulotto2008full}. 
Several designs based on all-metal or integrated meandered 1-D LWAs have been proposed \cite{MackeyMeanderedLWA,ScanningAritra,MeanderAritra} providing a frequency-scanning fan beam. However, 2-D configurations are required in order to achieve high-gain frequency-scanning pencil beams. 
In \cite{BeaskoetxeaAWPL15,KampouridouAccess21,shahbazian2024diminished}, radially corrugated planes have been studied, not requiring a dedicated beamforming network. Their rigorous design can be simplified by recurring to a local linearized model of the curved geometry (see, e.g., \cite{comite2020directive} and references therein). Unfortunately a single corrugated surface can suffer from direct source radiation in free space, leading to higher sidelobes and reduced maximum directivity. To suppress this contribution, the corrugated structure can be covered with a top metallic plate and radiate through azimuthal slots. The linearized model of this radial structure is therefore a parallel-plate waveguide (PPW) with corrugated bottom plate and periodic slits etched on its top plate (Fig. \ref{fig:geometry}).

\begin{figure*}[htbp]
    \centering

    \subfloat[]{%
        \includegraphics[width=0.32\textwidth]{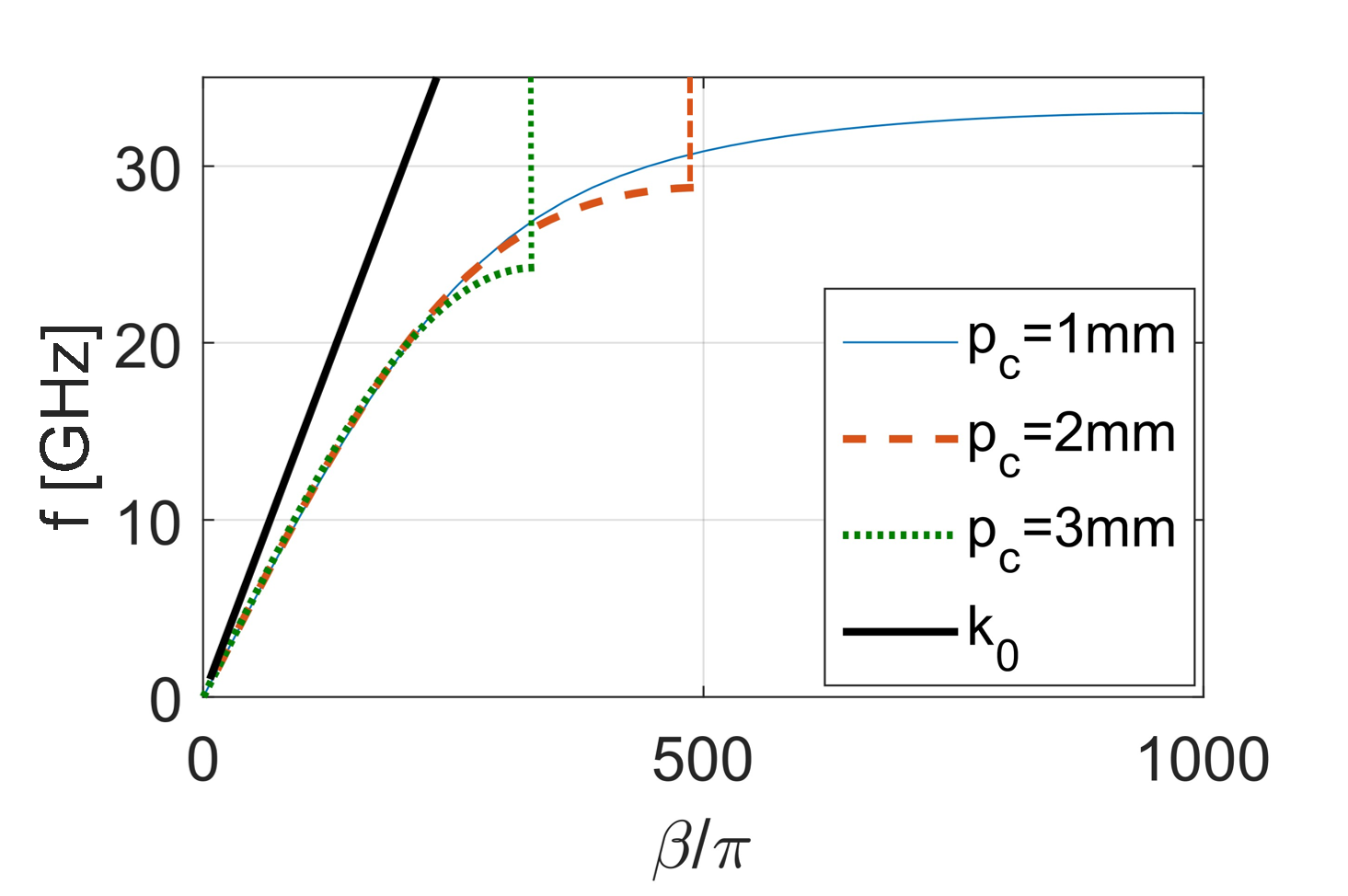}%
        \label{fig:beta_pc}
    }
    \hfill
    \subfloat[]{%
        \includegraphics[width=0.32\textwidth]{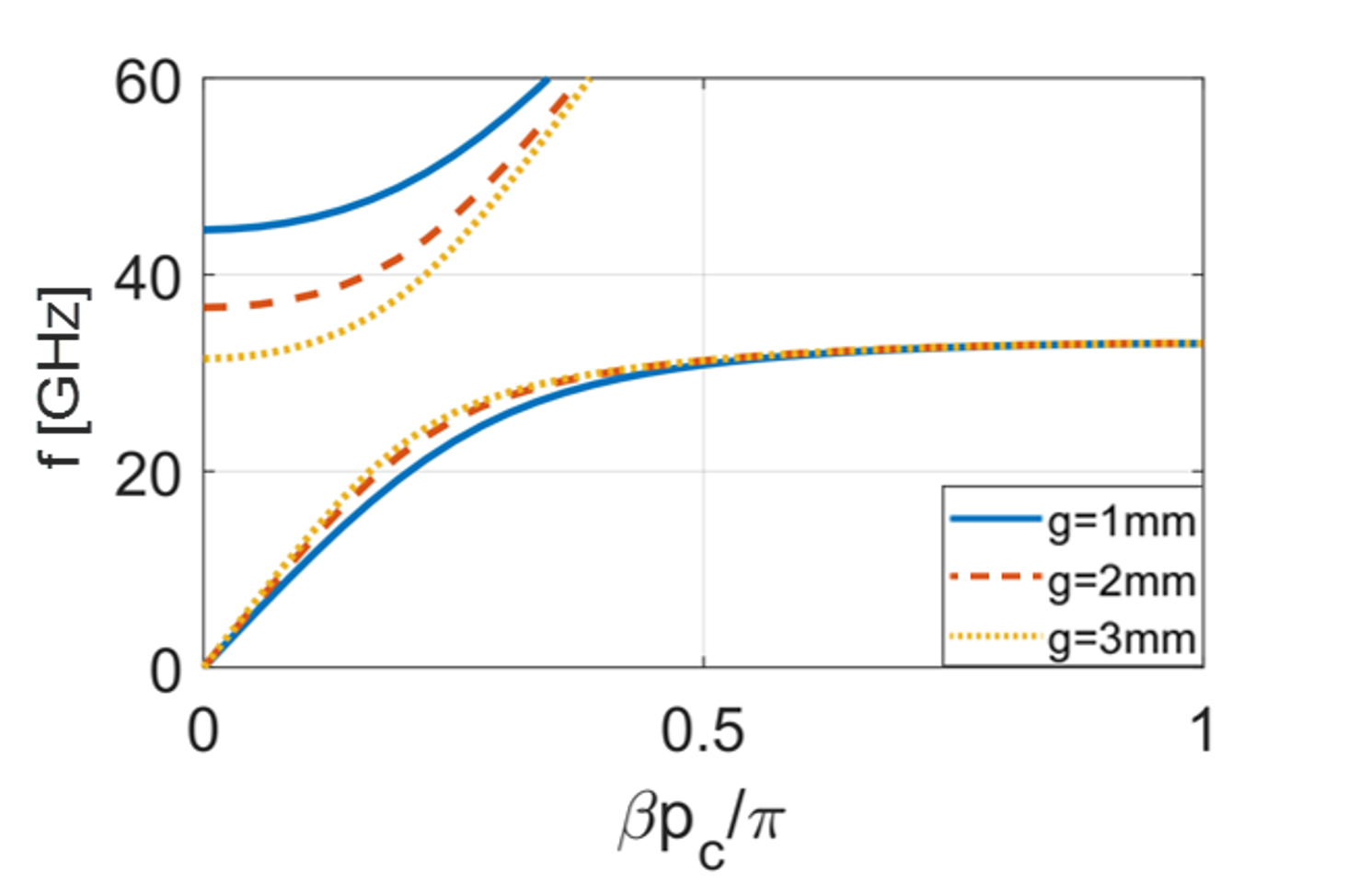}%
        \label{fig:beta_g}
    }
    \hfill
    \subfloat[]{%
        \includegraphics[width=0.32\textwidth]{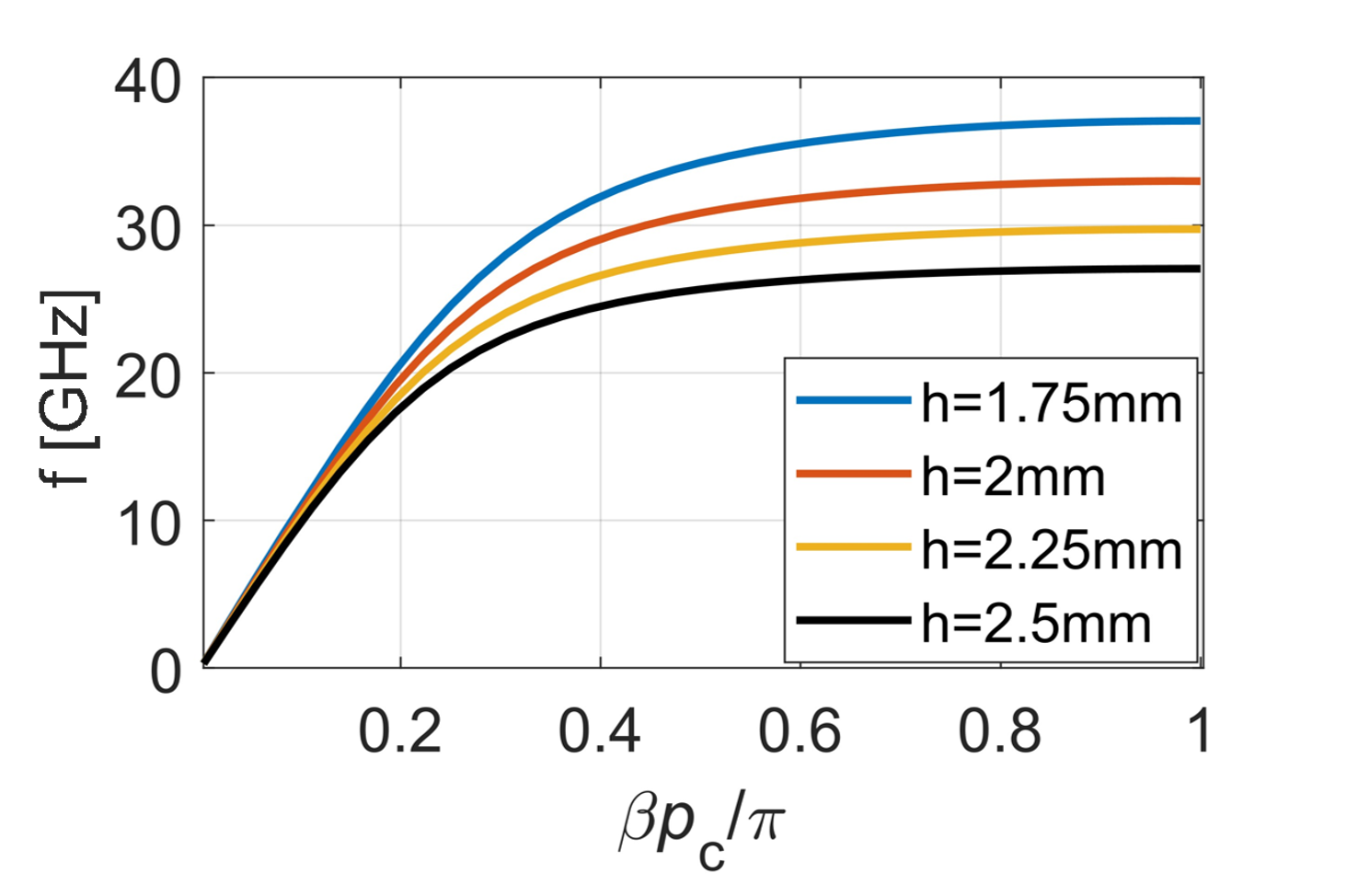}%
        \label{fig:beta_h}
    }

    \caption{Simulated $\beta$ of a closed unit cell with (a) different period of corrugation $p_\textrm{c}$, (b) different gap $g$ between two plates, (c) different corrugation depth $h$, with $p_c=1$ mm, $g=1$ mm.}
    \label{fig:beta_unit_cell}
\end{figure*}

This kind of structure has been studied in the absence of corrugation in \cite{Wu_Kishk}, where an interesting double feeding is proposed to overcome the presence of the OSB. However the lack of corrugation does not allow for a full control of frequency dispersion and the feeding configuration does not lead to a straightforward design of a radial structure. Other kinds of corrugated PPW, namely meandered PPW \cite{MeanderCTS,MeanderRidgedCTS}, have been developed in the last years as feeding networks capable to control the dispersion of Continuous Transverse Stubs (CTS)  \cite{MauroCTSKaBand,PotelonCTSEBand} and optimized to suppress their LWA OSB. These structures can reach high scanning velocity, but require a much higher profile than the one investigated here (at least 10 times thicker) and can be more subject to losses due to the long path traveled through the meanders.

We propose here, for the first time, the design of an all-metal low-profile LWA based on a corrugated PPW, supporting the radiation of a single fast spatial harmonic continuously scanning from the backward to the forward direction. In this configuration, the corrugations are designed to control the frequency dispersion of the guided wave, while the top plate can effectively shield the direct radiation of a source feeding the waveguide from the bottom plate. An interpretation of the role of geometrical parameters is given on the dispersion and radiation performance. The leaky-wave radiation is achieved through a carefully designed slot arrangement that suppresses the OSB. This approach also allows for studying different OSB transitions and their impact on radiation and matching issues. It will be shown that the diverse dispersive features of a well-known metal strip grating is found here, but avoiding the presence of lossy dielectrics.

The paper is structured as follows. Section \ref{sec:preliminary_desig} outlines the design principles for achieving a LWA with backward-to-forward transition at 30 GHz, and analyzes the impact of corrugation on its dispersive characteristics. Section \ref{sec:desig} presents the design capable to suppress the OSB on the structure. An efficient periodic Method-of-Moments (MoM) approach is employed for a rigorous computation of the phase and attenuation constant, leading to the final optimized design for OSB suppression. Conclusions are summarized in Section \ref{sec:concl}.

\begin{figure}
    \centering
    \includegraphics[width=0.4\textwidth]{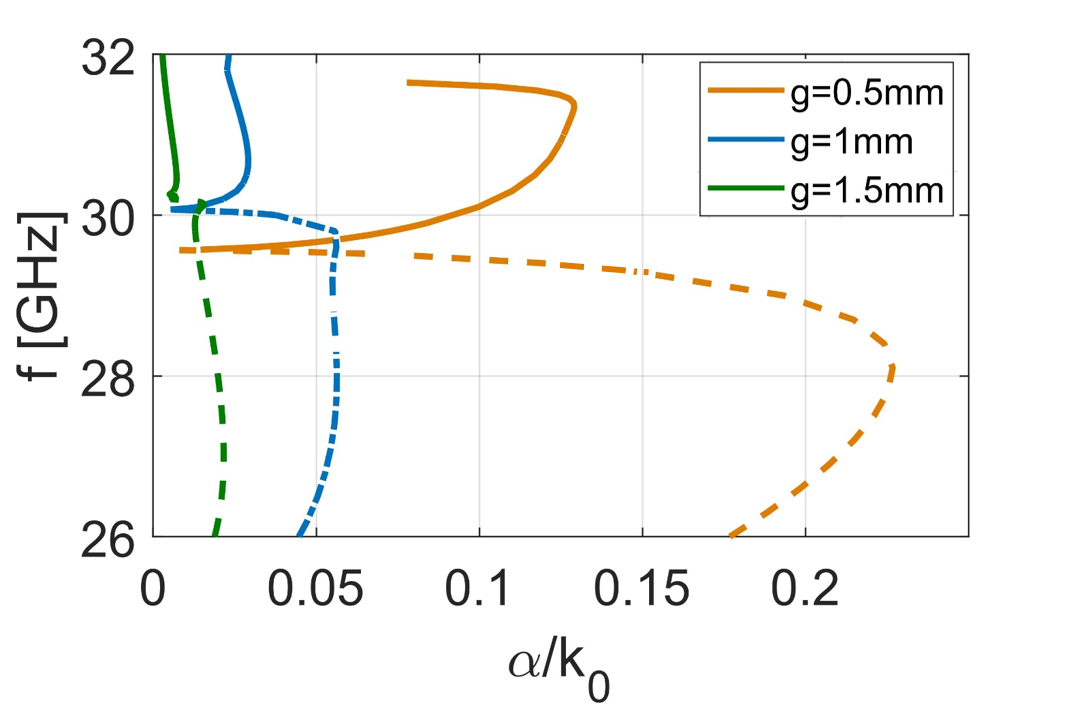} 
    \caption{Simulated $\alpha$ of a corrugated structure for various gaps $g$. Parameters: $h=2$ mm, $p_\text{c}=5$ mm, one slot ($w=2.5$ mm, $p_\text{s}=5$ mm).}
    \label{fig:alpha_g}
\end{figure}

\section{Design of corrugations and radiating slots}\label{sec:preliminary_desig}

Here the design process of the structure in Fig. \ref{fig:geometry} is described. The waveguide is rigorously 2-D (invariant along the $y$ direction) and we study here the high-frequency behavior of the TM$^x$ Floquet mode obtained by perturbing with slots a conventional corrugated waveguide, whose low-frequency guided-wave regime is well known \cite{elliott1954theory}. The modal complex wavenumber is $k_x=\beta - j\alpha$ of the wave propagating in the structure, where $\beta$ is the phase constant and $\alpha$ is the attenuation constant. The phase constant $\beta$ determines the main radiation direction $\theta$, which can be estimated using the relation $\theta=\sin^{-1}(\beta/k_0)$. Variations in $\beta$ with frequency directly influence the beam scanning behavior. The attenuation constant $\alpha$ accounts for the losses (in the case of a lossless structure, only those due to leakage \cite{paulotto2009novel}). It determines the antenna length for a given efficiency, and the beamwidth of the radiation pattern. 

The design consists of two main steps. First, the corrugation is analyzed to engineer its frequency dispersion. Then periodic slots are introduced to convert the guided-wave mode into a leaky regime. The aim is obtaining a broadside radiation around 30 GHz and a fast continuous scanning from backward to forward by suppressing the OSB. The remainder of this section presents the design methodology and simulation results of the closed corrugated PPW and the choice of the slot period. Slot radiation will be discussed in the next section.

Due to the periodic nature of the corrugations, the dispersion characteristics of the closed PPW can be analyzed by studying a single unit cell with a period of $p_\textrm{c}$. The phase constant obtained using the Eigenmode Solver in CST \cite{CST} is shown in Fig. \ref{fig:beta_unit_cell} for different corrugation periods $p_\textrm{c}$ and plate separation gaps $g$. The corrugation width $w_c$ is fixed at $p_\textrm{c}/2$, as its influence on the dispersion behavior is relatively moderate.

\begin{figure}
    \centering
    \subfloat[]{%
        \includegraphics[width=0.36\textwidth]{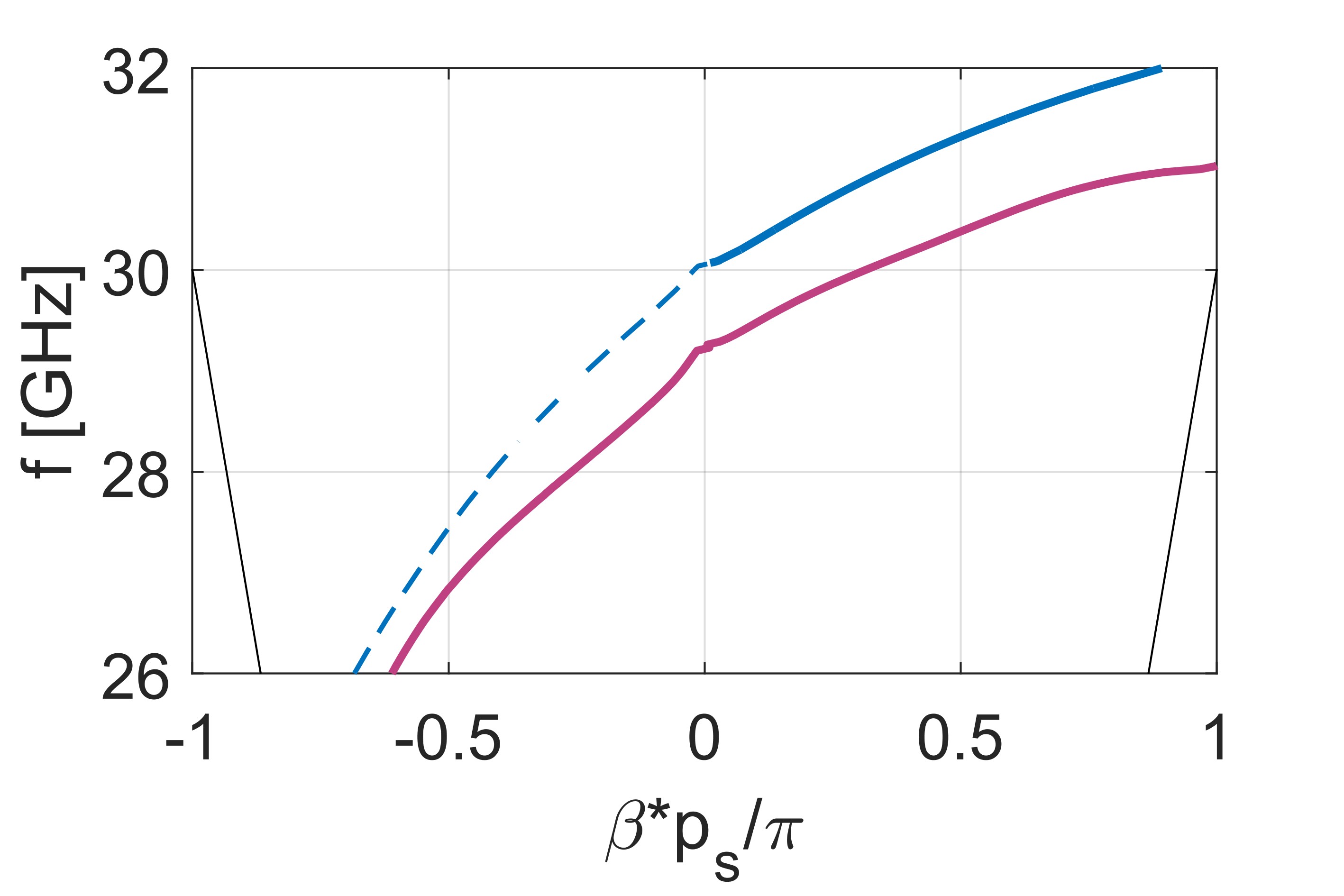}%
    }
    \hfill
    \subfloat[]{%
        \includegraphics[width=0.36\textwidth]{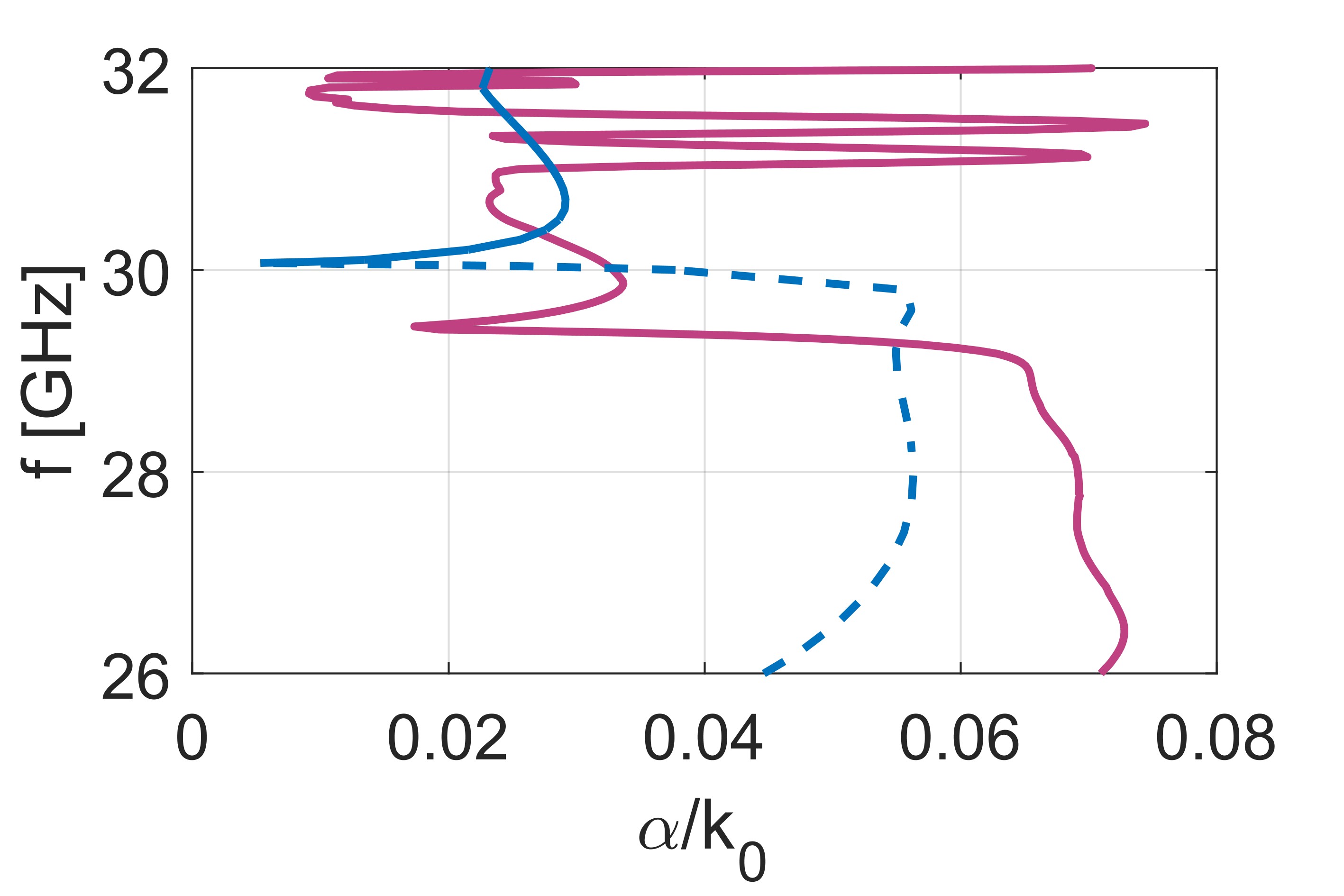}%
    }
    \caption{Simulated (a) $\beta$ and (b) $\alpha$ of a corrugated PPW with one slot $w=2.5$ mm periodically placed on the top plate. MMTMM results (\emph{purple lines}); MoM results (\emph{blue lines}), backward proper harmonic (\emph{dashed lines}), forward improper harmonic (\emph{solid lines}). Black lines in (a) represent the lines of light.}
    \label{fig:dip_1slot}
\end{figure}

Fig. \ref{fig:beta_unit_cell}(a) demonstrates that at low frequencies the phase constant of the fundamental mode remains nearly unchanged for varying values of $p_\textrm{c}$. This is coherent with the equivalent reactance of corrugations, which is only dependent on $w/p_\textrm{c}$ \cite{Marko_Zvonimir_Kildal_2010}. However, the region where the slow wave has a strong plasmonic behavior translates toward lower frequencies as $p_c$ increases. 
In Fig. \ref{fig:beta_unit_cell}(b) a larger gap $g$ between the corrugations and the top metallic plate results in the lowering of the cutoff frequency of the first higher-order mode. 
In order to ensure monomodal operation, a value of $g$ less than 2 mm is considered. Fig. \ref{fig:beta_unit_cell}(c) illustrates the dependence of $\beta$ on the corrugation depth $h$. A larger $h$ significantly enhances the dispersion around 30 GHz (the dispersion curve being closer to a horizontal line). This means that increasing the corrugation depth can effectively improve the scanning velocity of the LWA. However, if $h$ is too large the fundamental mode is in stopband at 30 GHz. To achieve broadside radiation at approximately $30$ GHz using the fundamental mode, a depth of $h$ less than $2.25$ mm is therefore considered.

\begin{figure}
\centering
    \subfloat[]{%
        \includegraphics[width=0.36\textwidth]{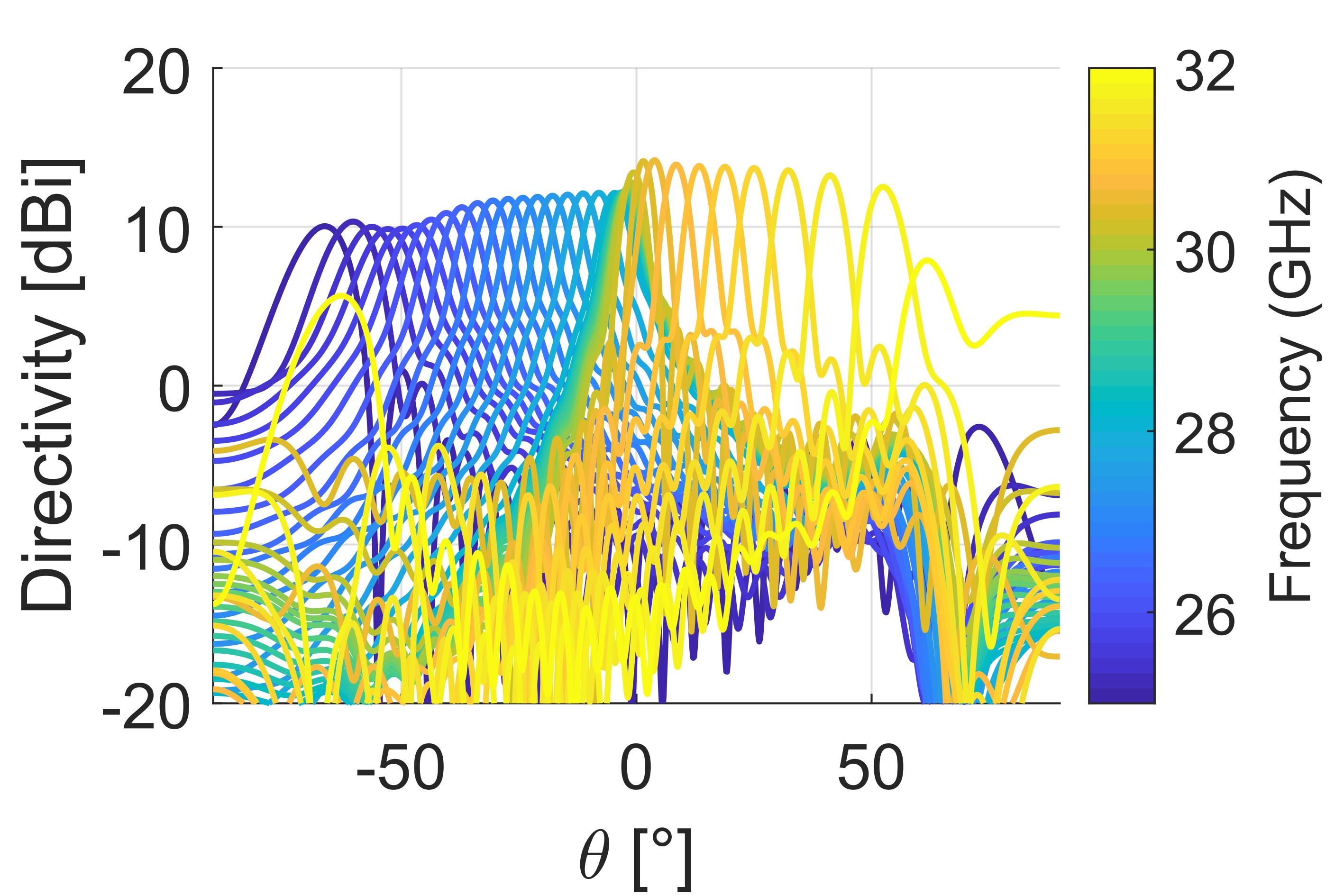}%
    }
    \hfill
    \subfloat[]{%
        \includegraphics[width=0.36\textwidth]{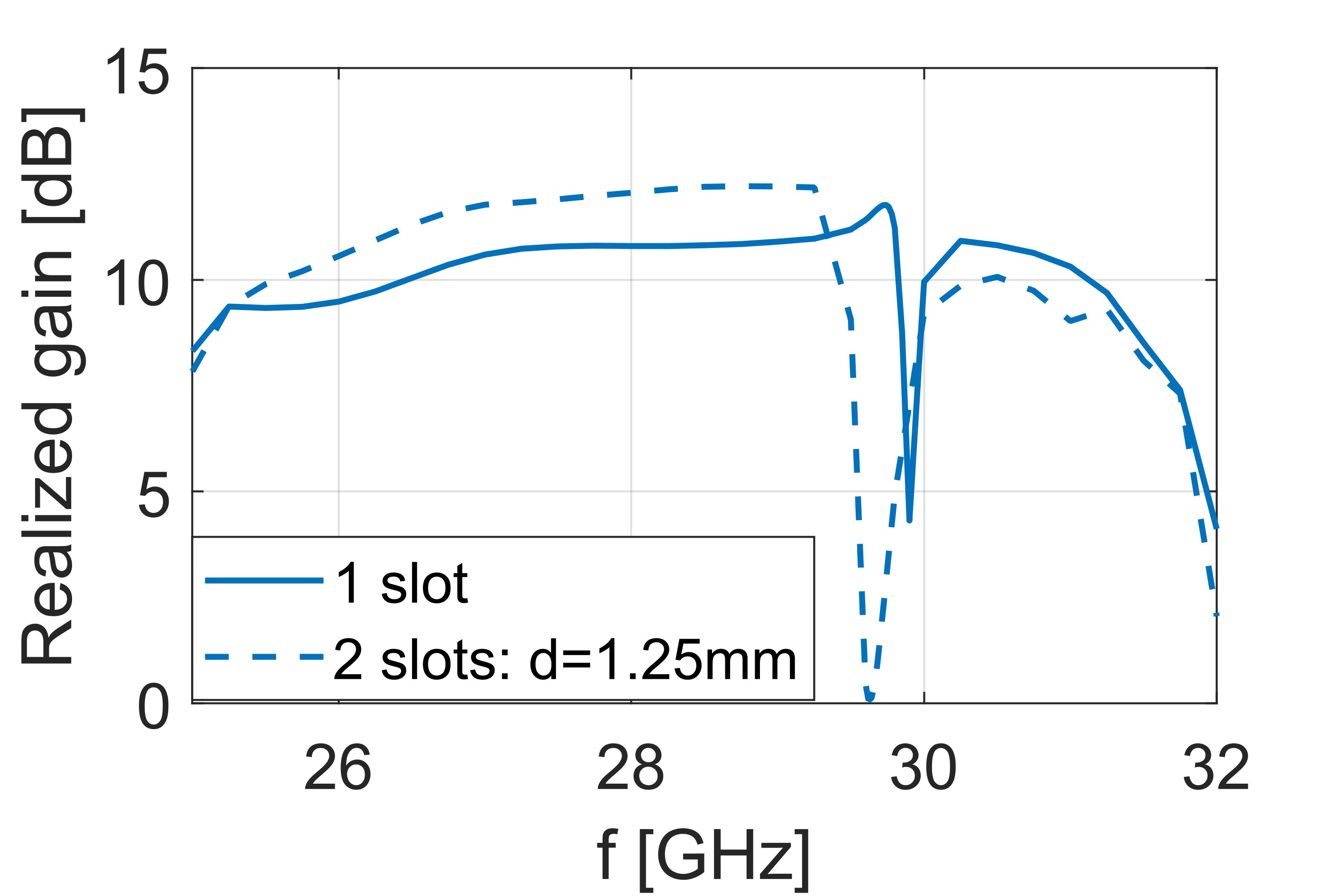}%
    }
\caption{(a) Simulated radiation pattern of a 20-cell corrugated PPW with one radiating slot having a width of $w=2.5$ mm. (b) Realized gain of a 20-cell corrugated PPW with one radiating slot ($w=2.5$ mm) and with two slots ($w_1=1.5$ mm, $w_2=0.1$ mm, mutual distance $d=1.25$ mm).}
\label{fig:radiation_1slot}
\end{figure}

To excite a fast harmonic contributing to radiation, periodic slots with period $p_{\rm \textrm{s}}$ on top of the PPW will be introduced \cite{volakis07AEHbook}. As first approximation, spatial harmonics created by the periodic slots are obtained as spectral translations, with spectral period $2\pi/p_{\rm \textrm{s}}$, of the dispersive curve of the closed-waveguide. To ensure broadside radiation at $30$ GHz using the $-1$ harmonic, the condition $\beta_{-1}=\beta_0 - \frac{2\pi}{p_\textrm{s}}=0$ must be satisfied at 30 GHz. This yields $p_\textrm{s}=2\pi/\beta_{0}=\lambda_\textrm{g}$, where $\beta_{0}$ is taken as the phase constant in the closed structure at 30 GHz, and $\lambda_g$ is the corresponding guided wavelength.

In addition to the OSB, periodic LWAs experience other stopbands due to the coupling between forward- and backward-traveling harmonics. Close to the operating frequency band, a stopband can arise between the forward $n=-1$ and the backward $n=-2$ harmonics. Furthermore, the presence of the $n=-2$ space harmonic in the visible region would lead to an undesired secondary radiation lobe. Extending the scanning range while ensuring that only the $n=-1$ space harmonic contributes to radiation enforces a constraint on the maximum value of $p_\textrm{s}$, which corresponds to a constraints on a minimum admissible $\beta_0$ at $30$ GHz (that can be controlled by tuning the corrugation depth $h$). Considering these factors, $h=2$ mm is selected, resulting in a slot period of $p_\textrm{s}=5$ mm. Higher $h$ would increase the frequency dispersion, but would make the working frequencies too close to the corrugation stopband, leading to difficult matching and high impact of metallic losses.

\section{Open stopband suppression}\label{sec:desig}
Having established the corrugation that provides the desired phase constant $\beta$ and the period of slots $p_s$ that creates a fast harmonic radiating at broadside at $30$ GHz, we now proceed to the second step: designing a waveguide with periodic radiating slots to achieve continuous beam scanning.

\subsection{Dispersive Analysis}

With the slot period determined, we now perform a rigorous dispersive analysis of the corrugated structure using an in-house numerical solver based on an efficient periodic Method of Moments (MoM) approach \cite{tong2024all}, among the few approaches allowing accurate characterization of the propagation and leakage behavior induced by the slots \cite{ComiteMoM,MoMPaper,Craeye_multisource,Coupled_Wave_Leaky}. It computes the complex wavenumber of the TM$^x$ Floquet mode by means of a periodic Electric-Field Integral Equation (EFIE) formulation. Approaches based on commercial solvers \cite{MMTMM-primer} avoid the development of an \textit{ad-hoc} code, but can suffer of inaccuracies when considering leaky waves, especially when the transition from backward to forward regime is analyzed and the leaky attenuation constant is of interest. In addition, an {\it ad-hoc} MoM approach allows for continuously following each and every identified modal solution and has the advantage of a simple control of the solution accuracy, mainly dependent on the meshing quality of the metallic lines.  

\begin{figure*}
    \centering

    \subfloat[]{%
        \includegraphics[width=0.35\textwidth]{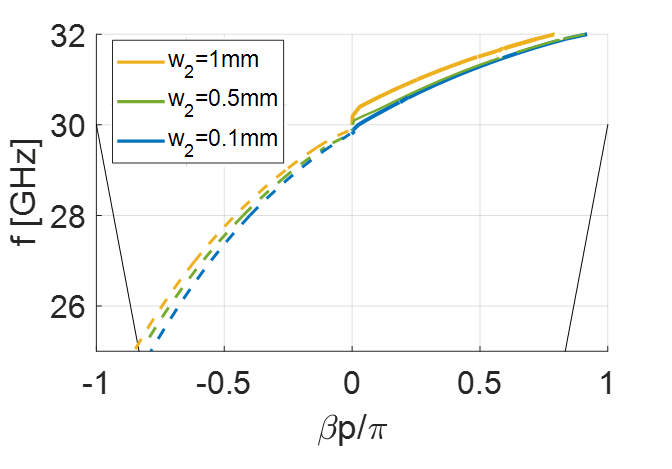}%
        \label{fig:beta_slot_pc}
    }
    \subfloat[]{%
        \includegraphics[width=0.35\textwidth]{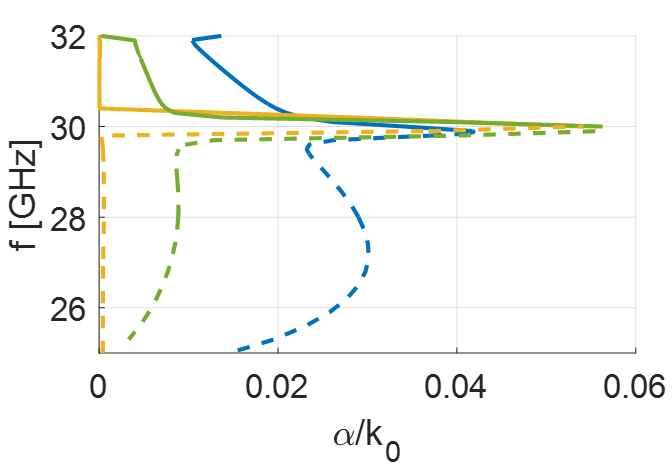}%
        \label{fig:alpha_slot_pc}
    }

    \caption{$\beta$ and $\alpha$ of a corrugated PPW with two slots periodically placed on the top plate. $w_1=1.5$ mm, inter slot distance $d=1.25$ mm. Backward proper harmonic (\emph{dashed lines}), forward improper harmonic (\emph{solid lines}).}
    \label{fig:dip_2slot(w2)}
\end{figure*}

\begin{figure*}
    \centering

    \subfloat[]{%
        \includegraphics[width=0.35\textwidth]{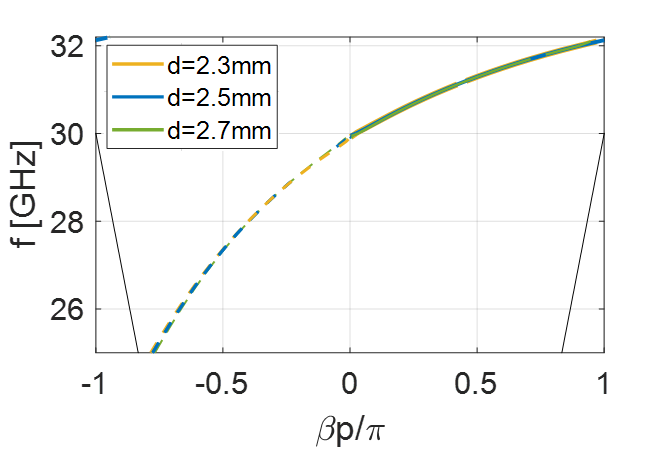}%
        \label{fig:beta_slot_g}
    }
    \subfloat[]{%
        \includegraphics[width=0.35\textwidth]{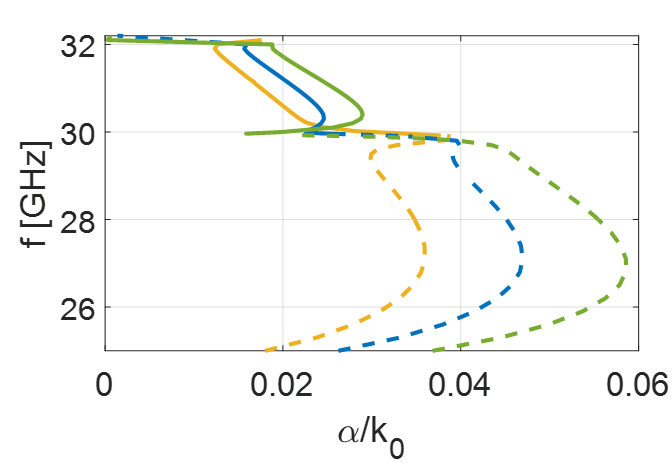}%
        \label{fig:alpha_slot_g}
    }

    \caption{$\beta$ and $\alpha$ of a corrugated PPW with two slots periodically placed on the top plate. $w_1=1.5$ mm, $w_2=0.1$ mm. Different values of inter-slot distance $d$. Backward proper harmonic (\emph{dashed lines}), forward improper harmonic (\emph{solid lines}).}
    \label{fig:dip_2slot(d)}
\end{figure*}

For a TM$^x$ polarization, the EFIE enforcing null electric field on metallic surfaces in absence of a source is \cite{Peterson}
\begin{equation}\label{equ:EFIE}
\hat{\textbf{n}}\times\int_C \left(1+\frac{1}{k^2}\nabla\nabla\cdot\right)\textbf{J}\left(\textbf{r}'\right)G_\text{p}\left[\textbf{r},\textbf{r}';k_{x}\left(\omega\right),\omega\right]\text{d}\textbf{r}'=\textbf{0}
\end{equation}
where $\textbf{J}\left(\textbf{r}\right)$ is the modal current flowing on the metallic line $C$ and $\hat{\textbf{n}}$ is the normal vector to $C$. $G_\text{p}$ is the periodic Green's function of the 2-D space with periodicity along the $x$ direction. The source and observation points on $C$ are $\textbf{r}'$ and $\textbf{r}$, respectively. The complex Floquet wavenumber $k_x$ is determined as a function of the angular frequency $\omega$ such than a non-null current can satisfy \eqref{equ:EFIE} with a null right-hand side (absence of sources).

\begin{figure*}
    \centering

    \subfloat[]{%
        \includegraphics[width=0.32\textwidth]{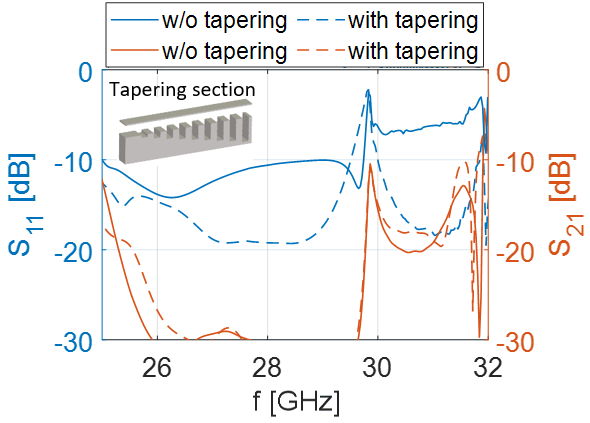}%
    }
    \hfill
    \subfloat[]{%
        \includegraphics[width=0.32\textwidth]{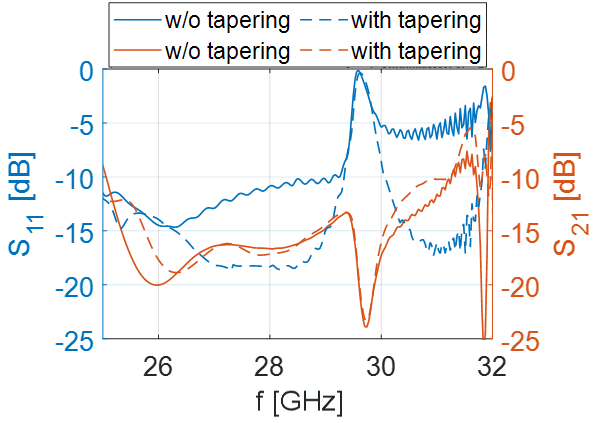}%
    }
    \hfill
    \subfloat[]{%
        \includegraphics[width=0.32\textwidth]{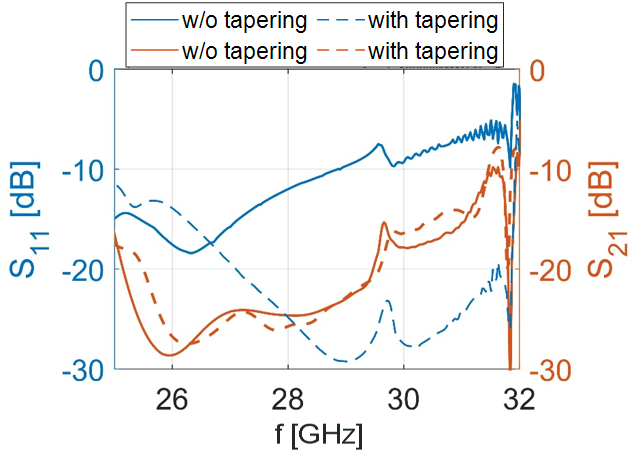}%
    }

    \caption{$S$-parameters of (a) structure with one slot $w_1 = 2.5$ mm, 
    (b) structure with two slots $w_1 = 1.5$ mm, $w_2 = 0.1$ mm, inter-slot distance $d = 1.25$ mm, and 
    (c) structure with two slots $w_1 = 1.5$ mm, $w_2 = 0.1$ mm, inter-slot distance $d = 2.5$ mm.}
    \label{fig:S_parameters}
\end{figure*}

\begin{figure}
\centering
    \subfloat[]{%
        \includegraphics[width=0.38\textwidth]{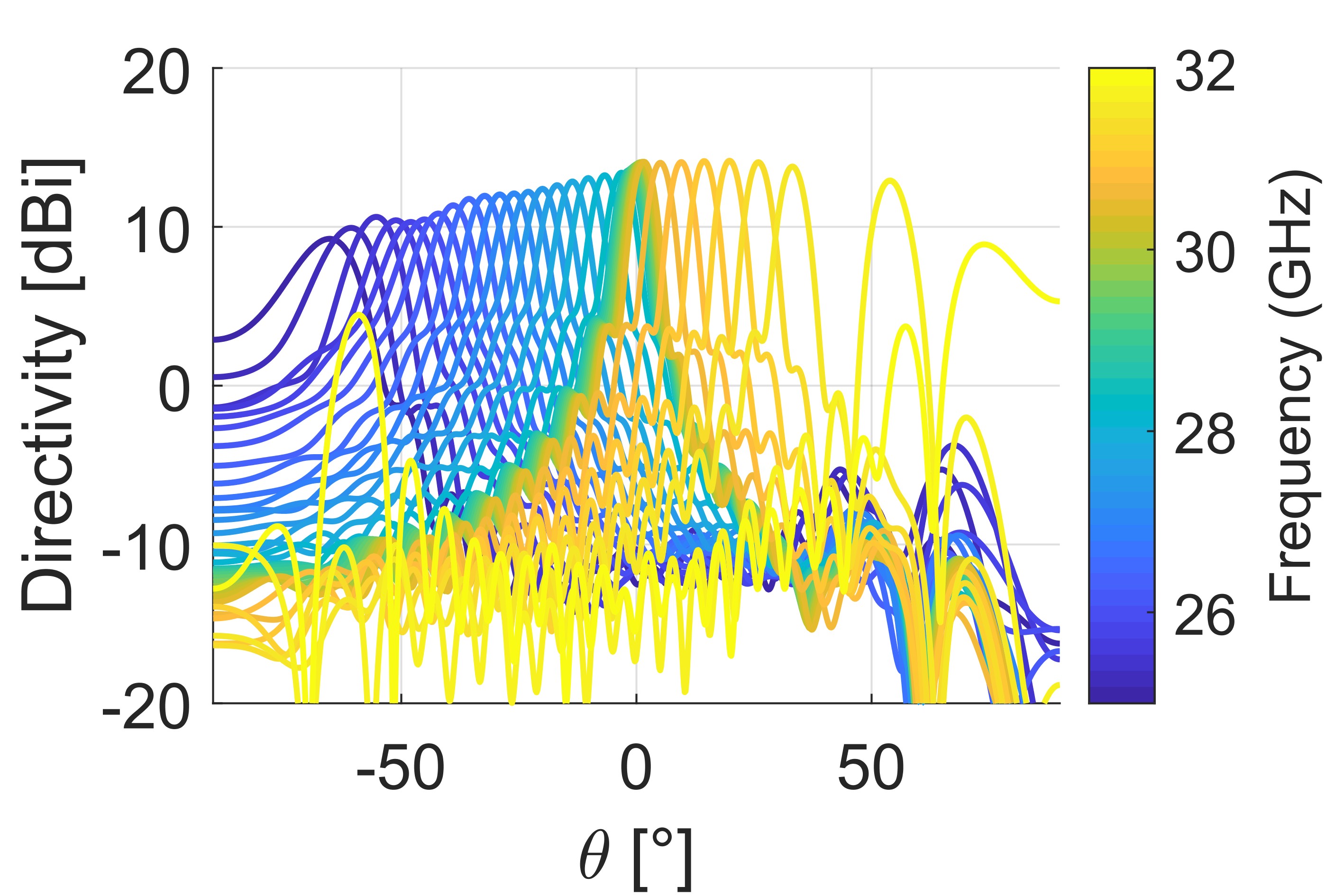}%
    }
    \hfill
    \subfloat[]{%
        \includegraphics[width=0.38\textwidth]{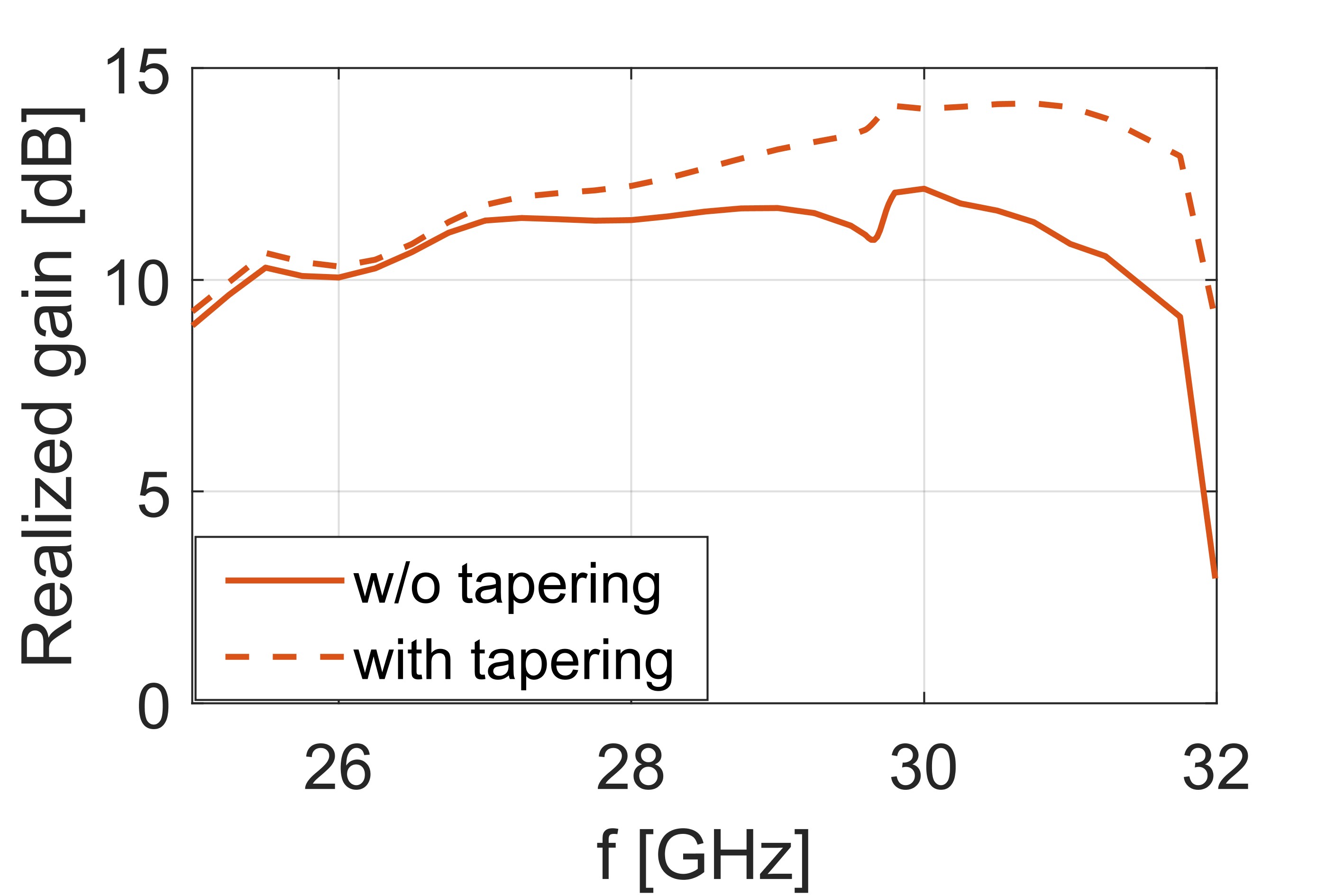}%
    }
    \hfill
    \subfloat[]{%
        \includegraphics[width=0.38\textwidth]{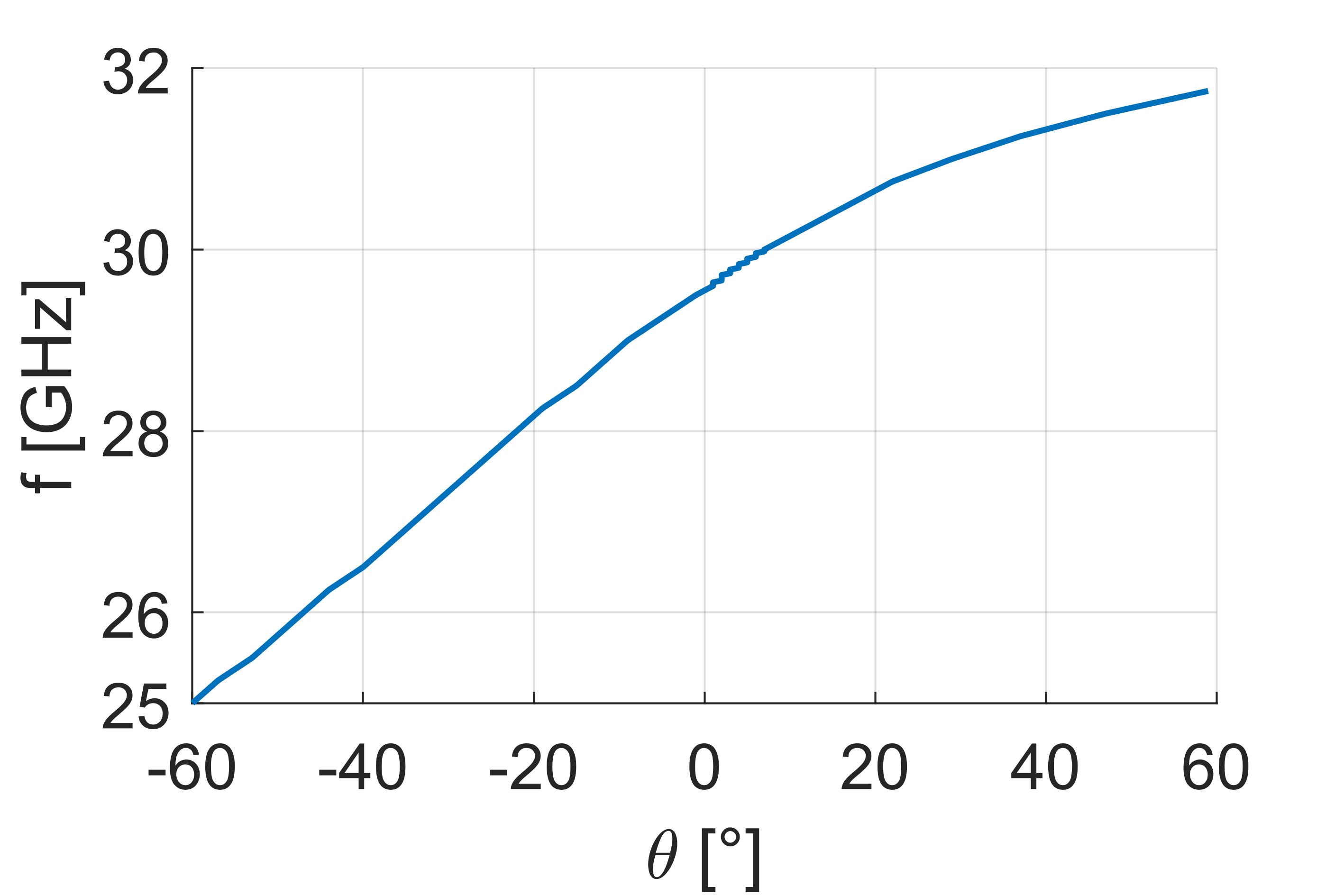}%
    }
\caption{(a) Radiation pattern, (b) maximum realized gain of a 20-cell corrugated with two radiating slots, (c) pointing angle of the beam; $w_1=1.5$ mm and $w_2=0.1$ mm, $d=2.5$ mm.}
\label{fig:radiation_2slot}
\end{figure}

The EFIE is conventionally discretized by means of a Galerkin MoM approach (piecewise linear basis and testing functions defined on sub-segments of the line $C$). The periodic Green's function $G_\text{p}$ is expressed as an Ewald sum, granting Gaussian convergence even in the presence of complex wavenumbers \cite{EwaldCapolino}, \cite{EwaldValerio}. The logarithmic singularities are extracted from the MoM integration and integrated in closed form \cite{tong2024all}.

The determinant of the MoM matrix is set to zero in order to look for complex solutions $k_x\left(\omega\right)$ of the dispersion problem, solved numerically with a Padé approximant \cite{zero_find_algo_Pinto}. Backward leaky harmonics ($\beta_{x_n}\alpha_{x_n}< 0$) are searched with a proper definition (their wavenumber along $z$ having a negative imaginary part, fulfilling the radiation condition), while forward leaky harmonics ($\beta_{x_n}\alpha_{x_n}>0$) are searched with an improper definition (their wavenumber along $z$ having a positive imaginary part, not fulfilling the radiation condition).

\subsection{Continuous scanning with OSB suppression}
In the open waveguide, the gap $g$ between the plate and the corrugations strongly influence the attenuation constant $\alpha$. To determine an appropriate value of $g$ that yields a moderate and controllable leakage rate, a parametric study is conducted using the MoM. As shown in Fig. \ref{fig:alpha_g}, $\alpha$ increases significantly as $g$ decreases. Based on this observation, $g=1$ mm is selected as a compromise that provides a moderate value of $\alpha$ suitable for subsequent slot optimization. To further validate these results, the dispersive behavior of this structure is also analyzed using the Multimodal Transmission Matrix Method (MMTMM) \cite{MMTMM-primer}. The MMTMM analysis is performed with a cascade of 3 unit cells and 2 background modes at each port. The results from both methods for the case of $g=1$ mm are presented in Fig. \ref{fig:dip_1slot}, showing good agreement. The slight disagreement is attributed to structural truncation and the absence of mutual coupling among adjacent cells in the MMTMM approach \cite{MMTMM-primer}.

The radiation pattern of a 20-cell cascade fed by an ideal PPW waveguide port and closed on a second identical waveguide port is shown in Fig. \ref{fig:radiation_1slot}(a), indicating a scanning range from $-63^\circ$ to $53^\circ$. To compute the directivity, the $E$-field distribution $E_x(x)$ on the top plate along the propagation direction ($x$-axis) is extracted from CST full-wave simulations.The bidimensional directivity $D(\theta)$ is
\begin{equation}\label{equ:Directivity}
D(\theta) = \frac{|F(\theta)|^2}{\frac{1}{\pi}\int_{-\frac{\pi}{2}}^\frac{\pi}{2}|F(\theta)|^2\text{d}\theta} 
\end{equation}
with
\begin{equation}\label{equ:radiation function}
F(\theta) = \int_L M_y(x) e^{j k_0 \sin\left(\theta\right) \, x} \, \text{d}x
\end{equation}
where $L$ is the aperture and $M_y(x)=2E_x(x)$ is the equivalent magnetic surface current density on it, an infinite PEC being assumed and removed by virtue of the image theorem.

The transition from backward to forward radiation occurs at $29.85$ GHz, very close to the frequency at which $\beta_{-1}=0$ is observed in the dispersion behavior calculated by the MoM (see Fig. \ref{fig:dip_1slot}(a)), further confirming MoM accuracy. As expected, an OSB appears when the attenuation constant $\alpha$ drops to zero at the frequency where $\beta_{-1}=0$ (see Fig. \ref{fig:dip_1slot}), due to the standing-wave effect caused by space harmonics coupling \cite{CollinZuck}. This results in a significant reduction of the realized gain \cite{paulotto2009novel}\cite{comite2018planar}, as shown by the solid curve in Fig. \ref{fig:radiation_1slot}(b). To suppress the OSB an asymmetry is introduced \cite{baccarelli2019open} in the form of a second slot of different width.

\begin{figure}
    \centering
    \includegraphics[width=0.4\textwidth]{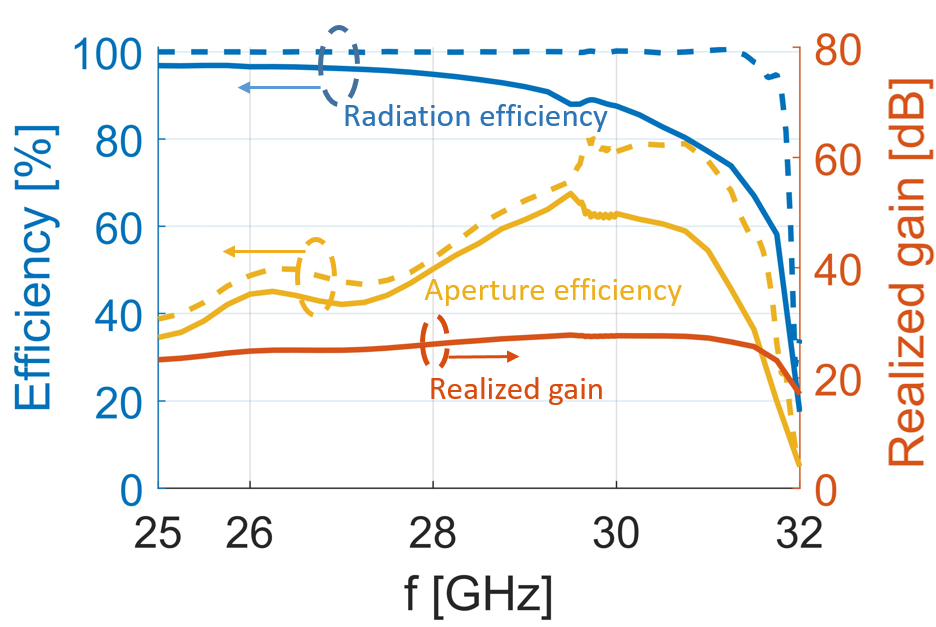} 
    \caption{Simulated radiation efficiency (\emph{blue lines}), aperture efficiency (\emph{yellow lines}) and realized gain (\emph{red line}) for a structure of width 71 mm. Lossless structure (\emph{dashed lines}) and lossy structure (\emph{solid lines}).}
    \label{fig:Efficiencies}
\end{figure}

The effects of various slot parameters, including the individual slot widths and the distance between them, on the antenna radiation performance are investigated. Initially, the inter-slot distance is chosen as $d=\frac{\lambda_{\rm g}}{4}=1.25$ mm to reduce the OSB \cite{baccarelli2019open}. The simulated $\beta$ and $\alpha$ for different values of $w_2$ are shown in Fig. \ref{fig:dip_2slot(w2)}, with a constant $w_1=1.5$ mm. In this new configuration $\alpha$ no more drops to zero; however, this does not indicate an efficient broadside radiation. Rather, the observed behavior of $\alpha$ resembles now to that of a closed stopband (CSB), where a drastic increase in $\alpha$ and a constant phase constant is observed in a frequency range. The large values of $\alpha$ in these cases are predominantly due to reactive effects, resulting in an input mismatch and a significantly reduced radiation efficiency \cite{paulotto2009novel}. This degradation is reflected in the realized gain, as shown by the dashed curve in Fig. \ref{fig:radiation_1slot}(b). The results in Fig. \ref{fig:dip_2slot(w2)} reveal that the stopband effect is mitigated for small values of $w_2$, namely when $w_2$ is decreased to $0.1$ mm. As $w_2$ is further reduced, one could see that the maximum value of $\alpha$ continues to decrease, albeit only slightly. Based on this finding, while maintaining $w_1=1.5$ mm and $w_2=0.1$ mm, we further adjusted the inter-slot distance $d$ in Fig. \ref{fig:dip_2slot(d)}. With $d=2.5 {\rm \,mm}=\lambda_{\rm g}/2$, $\alpha$ exhibits neither a drop to zero nor a pronounced peak, indicating that the reactive effect is effectively canceled and the OSB is suppressed. 


It should be noted that the optimal inter-slot distance observed here is $\lambda_{\rm g}/2$. This differs from the conventional $\lambda_{\rm g}/4$ spacing between radiating elements typically employed for OSB suppression \cite{paulotto2009novel,baccarelli2019open}. This difference can be attributed to the specific characteristics of the proposed configuration: the behavior of the two slots is here very different due to their extreme different width. It can be shown with \eqref{equ:Directivity} that the small slot is not significantly contributing to radiation, but rather canceling the reactive load of the larger slot at the OSB.  

To ensure good impedance matching when excited by a conventional waveguide port, a gradually corrugated tapering section is introduced at both ends of the structure \cite{elliott1954theory}. The simulated $S$-parameters of the structures with and without the tapering section for the three cases (OSB, CSB, and suppressed stopband) are shown in Fig. \ref{fig:S_parameters}. As observed in Figs. \ref{fig:S_parameters}(a) and (b), for both the OSB and CSB cases, the reflection coefficient within the stopband remains high even after introducing the tapering section, indicating that the impedance matching cannot be effectively improved with a simple tapering. In contrast, in the OSB-suppressed case (see Fig.~\ref{fig:S_parameters}(c)) the use of the tapering section leads to a significant reduction in the reflection coefficient, demonstrating successful impedance matching over the entire bandwidth of interest.

The resulting radiation patterns and the realized gain for a 20-cell configuration with the suppressed OSB are in Fig. \ref{fig:radiation_2slot} across the entire frequency range. They both exhibit broadside radiation at $29.7$ GHz and a backward-to-forward scanning from $-60^\circ$ to $50^\circ$ in the frequency range $[25,31]$ GHz (21.4\% relative bandwidth) [see Fig. \ref{fig:radiation_2slot}(c)]. Around $32$ GHz the $n=-2$ harmonic enters the visible region [see Fig. \ref{fig:dip_2slot(d)}], resulting in a secondary lobe and degrading the main beam quality. This scanning velocity is comparable to the one obtained with CTS fed by meanders having a thickness ten times larger then the corrugation depth chosen here \cite{MeanderCTS}.

The effect of copper losses is studied by making radiate in CST the LWA in a three-dimensional space. Its length along the $y$ direction is 71 mm (close to the antennas realized in \cite{MeanderCTS,MeanderRidgedCTS}). For consistency to previous results, the invariance along $y$ of the leaky mode is kept by enforcing $xz$ PMC planes a TEM-mode feeding. In Fig. \ref{fig:Efficiencies} the maximum gain, radiation and aperture efficiencies are shown, to be considered upper bounds of the values expected with a realistic feeder as in \cite{MeanderCTS,MeanderRidgedCTS}. Notice that the aperture efficiency achieved here is higher than common 2-D LWAs, due to the uniform excitation of a 1-D leaky mode by a source defined across the entire $zy$ plane. The LWA presented here achieves a stable gain and performances comparable to meander-fed CTS whose thicknesses are ten times larger than the structure proposed here.

\section{Conclusion}\label{sec:concl}
A corrugated all-metal PPW with periodic slots was investigated to design a LWA achieving beam scanning in the Ka-band with OSB suppression. 
A MoM-based dispersive analysis was conducted to accurately characterize the dispersion properties of structure. By carefully selecting the corrugation depth and corrugation and slot periodicity, a LWA scanning from $-60^\circ$ to $50^\circ$ in the Ka-band was realized, with broadside radiation at $29.7$ GHz. By introducing a suitably optimized second asymmetric slot, OSB suppression was successfully achieved, thus allowing for a continuous transition from backward to forward region.


\bibliographystyle{IEEEtran}
\balance
\bibliography{Leaky_mode_corrugated_PPW}

\begin{thebibliography}{10}
\providecommand{\url}[1]{#1}
\csname url@samestyle\endcsname
\providecommand{\newblock}{\relax}
\providecommand{\bibinfo}[2]{#2}
\providecommand{\BIBentrySTDinterwordspacing}{\spaceskip=0pt\relax}
\providecommand{\BIBentryALTinterwordstretchfactor}{4}
\providecommand{\BIBentryALTinterwordspacing}{\spaceskip=\fontdimen2\font plus
\BIBentryALTinterwordstretchfactor\fontdimen3\font minus
  \fontdimen4\font\relax}
\providecommand{\BIBforeignlanguage}[2]{{%
\expandafter\ifx\csname l@#1\endcsname\relax
\typeout{** WARNING: IEEEtran.bst: No hyphenation pattern has been}%
\typeout{** loaded for the language `#1'. Using the pattern for}%
\typeout{** the default language instead.}%
\else
\language=\csname l@#1\endcsname
\fi
#2}}
\providecommand{\BIBdecl}{\relax}
\BIBdecl

\bibitem{lenses_comm}
A.~Algaba-Brazález, P.~Castillo-Tapia, M.~C. Viganó, and O.~Quevedo-Teruel,
  ``Lenses combined with array antennas for the next generation of terrestrial
  and satellite communication systems,'' \emph{IEEE Communications Mag.},
  vol.~62, no.~9, pp. 176--182, 2024.

\bibitem{jackson2012leaky}
D.~R. Jackson, C.~Caloz, and T.~Itoh, ``Leaky-wave antennas,'' \emph{Proc.
  IEEE}, vol. 100, no.~7, pp. 2194--2206, 2012.

\bibitem{DirFindingJose}
A.~Gil-Martínez, M.~Poveda-García, J.~García-Fernández, M.~M. Campo-Valera,
  D.~Cañete-Rebenaque, and J.~L.~G. Tornero, ``Direction finding of {RFID}
  tags in {UHF} band using a passive beam-scanning leaky-wave antenna,''
  \emph{IEEE J. Radio Frequency Identification}, vol.~6, pp. 552--563, 2022.

\bibitem{paulotto2008full}
S.~Paulotto, P.~Baccarelli, F.~Frezza, and D.~R. Jackson, ``Full-wave modal
  dispersion analysis and broadside optimization for a class of microstrip
  {CRLH} leaky-wave antennas,'' \emph{IEEE Trans. Microwave Theory Tech.},
  vol.~56, no.~12, pp. 2826--2837, 2008.

\bibitem{MackeyMeanderedLWA}
A.~J. Mackay and G.~V. Eleftheriades, ``Meandered and dispersion-enhanced
  planar leaky-wave antenna with fast beam scanning,'' \emph{IEEE Antennas
  Wireless Prop. Lett.}, vol.~20, no.~8, pp. 1596--1600, 2021.

\bibitem{ScanningAritra}
A.~Roy, G.~Valerio, and J.~Sarrazin, ``A study on efficiency and frequency
  scanning in corrugated-waveguide-based leaky-wave antennas,'' in \emph{Proc.
  European Microwave Conf. (EuMC)}, 2024, pp. 445--448.

\bibitem{MeanderAritra}
------, ``Meander leaky wave antenna with open stop-band suppression for
  millimetre wave direction-finding,'' in \emph{Proc. European Conf. Antennas
  Prop. (EuCAP)}, 2025, pp. 1--5.

\bibitem{BeaskoetxeaAWPL15}
U.~Beaskoetxea, V.~Pacheco-Peña, B.~Orazbayev, T.~Akalin, S.~Maci,
  M.~Navarro-Cía, and M.~Beruete, ``77-{GHz} high-gain bull’s-eye antenna
  with sinusoidal profile,'' \emph{IEEE Antennas Wireless Propag. Lett.},
  vol.~14, pp. 205--208, 2015.

\bibitem{KampouridouAccess21}
D.~Kampouridou and A.~Feresidis, ``Broadband {THz} corrugated bull’s eye
  antennas,'' \emph{IEEE Access}, vol.~9, pp. 104\,460--104\,468, 2021.

\bibitem{shahbazian2024diminished}
C.~Shahbazian and A.~Kishk, ``The diminished edge diffraction effect bull's eye
  antenna,'' in \emph{Proc. Europ. Conf. Antennas Propag. (EuCAP)}, 2024, pp.
  1--4.

\bibitem{comite2020directive}
D.~Comite, M.~Kuznetcov, V.~G.-G. Buend{\'\i}a, S.~K. Podilchak, P.~Baccarelli,
  P.~Burghignoli, and A.~Galli, ``Directive {2-D} beam steering by means of a
  multiport radially periodic leaky-wave antenna,'' \emph{IEEE Trans. Antennas
  Propag.}, vol.~69, no.~5, pp. 2494--2506, 2020.

\bibitem{Wu_Kishk}
N.~Bayat-Makou, K.~Wu, and A.~A. Kishk, ``Single-layer substrate-integrated
  broadside leaky long-slot array antennas with embedded reflectors for {5G}
  systems,'' \emph{IEEE Trans. Antennas Propag.}, vol.~67, no.~12, pp.
  7331--7339, 2019.

\bibitem{MeanderCTS}
Y.~You, Y.~Lu, Y.~Wang, J.~Xu, J.~Huang, and W.~Hong, ``Enhanced pencil-beam
  scanning {CTS} leaky-wave antenna based on meander delay line,'' \emph{IEEE
  Antennas Wireless Prop. Lett.}, vol.~20, no.~9, pp. 1760--1764, 2021.

\bibitem{MeanderRidgedCTS}
Q.-D. Cao, X.-X. Yang, F.~Yu, and S.~Gao, ``High scanning rate millimeter-wave
  circularly polarized {CTS} leaky wave antenna,'' \emph{IEEE Trans. Antennas
  Propag.}, vol.~72, no.~7, pp. 6087--6092, 2024.

\bibitem{MauroCTSKaBand}
M.~Ettorre, F.~Foglia~Manzillo, M.~Casaletti, R.~Sauleau, L.~Le~Coq, and
  N.~Capet, ``Continuous transverse stub array for {Ka-Band} applications,''
  \emph{IEEE Trans. Antennas Propag.}, vol.~63, no.~11, pp. 4792--4800, 2015.

\bibitem{PotelonCTSEBand}
T.~Potelon, M.~Ettorre, L.~Le~Coq, T.~Bateman, J.~Francey, D.~Lelaidier,
  E.~Seguenot, F.~Devillers, and R.~Sauleau, ``A low-profile broadband 32-slot
  continuous transverse stub array for backhaul applications in {E-Band},''
  \emph{IEEE Trans. Antennas Propag.}, vol.~65, no.~12, pp. 6307--6316, 2017.

\bibitem{elliott1954theory}
R.~Elliott, ``On the theory of corrugated plane surfaces,'' \emph{IRE Trans.
  Antennas Propag.}, vol.~2, no.~2, pp. 71--81, 1954.

\bibitem{paulotto2009novel}
S.~Paulotto, P.~Baccarelli, F.~Frezza, and D.~R. Jackson, ``A novel technique
  for open-stopband suppression in {1-D} periodic printed leaky-wave
  antennas,'' \emph{IEEE Trans. Antennas Propag.}, vol.~57, no.~7, pp.
  1894--1906, 2009.

\bibitem{CST}
``{CST Microwave Studio},'' \url{http://www.cst.com/}, version: 2016.

\bibitem{Marko_Zvonimir_Kildal_2010}
M.~Bosiljevac, Z.~Sipus, and P.~S. Kildal, ``Construction of {G}reen's
  functions of parallel plates with periodic texture with application to gap
  waveguides - a plane-wave spectral-domain approach,'' \emph{IET Microwaves,
  Antennas Propagation}, vol.~4, no.~11, pp. 1799--1810, Nov. 2010.

\bibitem{volakis07AEHbook}
J.~L. Volakis, \emph{Antenna engineering handbook}.\hskip 1em plus 0.5em minus
  0.4em\relax New York, NY: McGraw Hill, 2007, 4th Ed.

\bibitem{tong2024all}
Y.~Tong, G.~Valerio, B.~Ambrogi, D.~Comite \emph{et~al.}, ``All-metal
  glide-symmetric slotted planar antennas: modal analysis,'' in \emph{Proc.
  European Conf. Antennas Propag. (EuCAP)}, 2024.

\bibitem{ComiteMoM}
D.~Zhang, D.~Comite, P.~Baccarelli, X.~Deng, X.~Zheng, A.~Galli, and
  P.~Burghignoli, ``Method-of-moments solver for all-metal corrugated
  structures: Leaky-mode analysis and open-stopband suppression,'' \emph{IEEE
  Trans. Microwave Theory Tech.}, pp. 1--10, 2023.

\bibitem{MoMPaper}
M.~Petek, J.~Rivero, J.~A. Vasquez-Tobon, G.~Valerio, O.~Quevedo-Teruel, and
  F.~Vipiana, ``{Method of Moments for the Dispersion Modelling of Glide-
  Symmetric Periodic Structures},'' \emph{IEEE Trans. Antennas Propag.}, in
  press.

\bibitem{Craeye_multisource}
M.~Bodehou, A.~Guth, K.~A. Khalifeh, D.~Heberling, and C.~Craeye, ``Multi-feed
  metasurface antennas: Direct numerical design and experimental validations,''
  \emph{IEEE Access}, vol.~11, pp. 35\,754--35\,762, 2023.

\bibitem{Coupled_Wave_Leaky}
D.~TomiĆ, F.~Mesa, and Z.~Šipuš, ``Glide-shifted dielectric gratings for
  independent leakage and phase tuning in leaky-wave antennas,'' \emph{IEEE
  Open J. Antennas Prop.}, pp. 1--1, 2025.

\bibitem{MMTMM-primer}
F.~Mesa, G.~Valerio, R.~Rodriguez-Berral, and O.~Quevedo-Teruel,
  ``{Simulation-Assisted Efficient Computation of the Dispersion Diagram of
  Periodic Structures: A comprehensive overview with applications to filters,
  leaky-wave antennas and metasurfaces},'' \emph{IEEE Antennas Propag. Mag.},
  vol.~63, no.~5, pp. 33--45, 2020.

\bibitem{Peterson}
F.~Peterson, S.~L. Ray, and R.~Mittra, \emph{Computational Methods for
  Electromagnetics}.\hskip 1em plus 0.5em minus 0.4em\relax Wiley-IEEE Press,
  1998.

\bibitem{EwaldCapolino}
F.~Capolino, D.~Wilton, and W.~Johnson, ``Efficient computation of the {2}-{D}
  {G}reen's function for 1-{D} periodic structures using the ewald method,''
  \emph{IEEE Trans. Antennas Propag.}, vol.~53, no.~9, pp. 2977--2984, 2005.

\bibitem{EwaldValerio}
G.~Valerio, P.~Baccarelli, P.~Burghignoli, and A.~Galli, ``Comparative analysis
  of acceleration techniques for 2-{D} and 3-{D} {G}reen's functions in
  periodic structures along one and two directions,'' \emph{IEEE Trans.
  Antennas Propag.}, vol.~55, no.~6, pp. 1630--1643, 2007.

\bibitem{zero_find_algo_Pinto}
V.~Galdi and I.~M. Pinto, ``A simple algorithm for accurate location of
  leaky-wave poles for grounded inhomogeneous dielectric slabs,''
  \emph{Microwave Optical Tech. Lett.}, vol.~24, no.~2, pp. 135--140, 2000.

\bibitem{CollinZuck}
R.~Collin and F.~Zucker, \emph{Antenna Theory}, ser. Inter-University
  Electronics Series.\hskip 1em plus 0.5em minus 0.4em\relax McGraw-Hill, 1969,
  no. pt. 2.

\bibitem{comite2018planar}
D.~Comite, V.~G.-G. Buend{\'\i}a, S.~K. Podilchak, D.~Di~Ruscio, P.~Baccarelli,
  P.~Burghignoli, and A.~Galli, ``Planar antenna design for omnidirectional
  conical radiation through cylindrical leaky waves,'' \emph{IEEE Antennas
  Wireless Propag. Lett.}, vol.~17, no.~10, pp. 1837--1841, 2018.

\bibitem{baccarelli2019open}
P.~Baccarelli, P.~Burghignoli, D.~Comite, W.~Fuscaldo, and A.~Galli,
  ``Open-stopband suppression via double asymmetric discontinuities in 1-{D}
  periodic 2-{D} leaky-wave structures,'' \emph{IEEE Antennas Wireless Propag.
  Lett.}, vol.~18, no.~10, pp. 2066--2070, 2019.

\end{thebibliography}

\end{document}